\documentstyle[12pt]{article}
\begin{document}
\newcommand{\be}{\begin{equation}}
\newcommand{\ee}{\end{equation}}
\newcommand{\bea}{\begin{eqnarray}}
\newcommand{\eea}{\end{eqnarray}}
\baselineskip=18pt
\begin{center}
{\large{\bf 
Fermion Pairing Dynamics in the Relativistic Scalar Plasma 
}}
\end{center}
\vspace{0.3cm}

\begin{center}
E. R. Takano Natti \footnote{Present address : Universidade do Norte 
do Paran\'a, Av. Paris, 675, CEP 86041-140, Londrina, Paran\'a, 
Brasil}, Chi-Yong Lin and A. F. R. de Toledo Piza\\ 
\end{center}

\begin{center}
{\it
Instituto de F\'{\i}sica, Universidade de S\~ao Paulo,\\  
Caixa Postal 66318, CEP 05389-970, S\~ao Paulo, S\~ao Paulo, Brazil}
\end{center}
\vskip 0.3cm
\begin{center}
P. L. Natti
\end{center}
\begin{center}
{\it
Departamento de Matem\'atica, Universidade Estadual de Londrina,\\ 
Caixa Postal 6001, Cep 86051-970, Londrina, Paran\'a, Brazil}
\end{center}

\baselineskip=24pt
\vskip 0.3cm
\begin{center}
{\bf ABSTRACT}
\end{center}
\vskip 0.1cm

\indent 
Using many-body techniques we obtain the time-dependent Gaussian
approximation for interacting fermion-scalar field models.
This method is applied to  an uniform system of
relativistic spin-$1/2$ fermion field coupled, through a Yukawa term, to
a scalar field in 3+1 dimensions, the so-called quantum scalar plasma
model. 
The renormalization for the resulting Gaussian mean-field equations,
both static and dynamical, are examined and initial conditions
discussed. 
We also investigate solutions for the gap equation and show that the
energy density has a single minimum. 

\vskip 0.3cm
\begin{center}
{\it Submitted to Phys. Rev. D15} 
\hspace{0.5cm} ({\it Revised Version}) 
\end{center}

\vskip 0.3cm

\vfill

\noindent PACS number(s): 11.10-z, 11.10.Lm, 11.10Gh, 21.60.Jz.
\hspace{\fill}


\newpage

\normalsize
\newpage
\baselineskip=23pt
\setcounter{equation}{0}
\begin{center}
{\bf I. Introduction}
\end{center}

\indent
The self-consistent mean-field method remains being one of the few
analytical tools available to investigate a variety of problems in
quantum field theory which cannot be handled in  perturbative theory.
These include static problems in the context of spontaneous symmetry
breaking \cite{St85}
as well as time-dependent problems involving the dynamics of
the inflation-driving scalar field in the early Universe \cite{Bo97}
and the
dynamics of relaxation process in the ultra-relativistic heavy-ion
experiments \cite{CBL97}. 
In its most used form the approximation is implemented, in the case of
boson fields, through of a time-dependent or independent  variational
principle using a Gaussian trial wavefunction in the Schroedinger
picture \cite{EPJ88}. 
However, systematic corrections to this approximation are still an open
problem.
Furthermore, for the case of fermion fields, the trial wavefunction
is not straightforward to take in a Gaussian form \cite{KSY90}.

\indent
On the other hand, the mean-field description of systems of interest
can be  obtained from the Heisenberg equation
\be
\label{1a}
i \langle \dot{\cal O}\rangle=Tr [{\cal O},H] {\cal F}_{0}\;\;,
\ee

\noindent
where ${\cal O}$ is a one-body operator and $H$ is the Hamiltonian of
the system. This can be kept under direct control when one takes a
gaussian-like density.
In a  formulation appropriate for the many-body theory we have 
\be
\label{1b}
{\cal F}_{0} =\frac{\exp \left[\sum_{(i,j)}
A_{i,j}\eta^{\dag}_{i}\eta_{j}+ B_{i,j}\eta^{\dag}_{i}\eta^{\dag}_{j}+
C_{i,j}\eta_{i}\eta_{j}\right]} {Tr \left\{\exp \left[\sum_{(i,j)}
A_{i,j}\eta^{\dag}_{i}\eta_{j}+ B_{i,j}\eta^{\dag}_{i}\eta^{\dag}_{j}+
C_{i,j}\eta_{i}\eta_{j}\right]\right\}} \;\;.
\ee 

\noindent
In (\ref{1b}) the
$\eta^{\dag}_{i}$ ($\eta_{i}$) are boson or fermion creation
(annhihilation) operators of a particle in orbital $i$ and 
${\cal F}_{0}$ can
be, of course, rewritten in a diagonal form when one uses the
Bogoliubov quasiparticle operators (cf. Eqs.(\ref{4a}) and (\ref{5a})). 
It is not difficult to see that this scheme corresponds to the usual
gaussian variational approximation in the functional Schroedinger
picture. 
There are, however, some important differences. 
For instance, the usual technique of nonrelativistic many-fermion
problem  can be readily extended to the Dirac field theory \cite{CS81}.
Thus, instead of using anticommuting Grassmann variables to write
Gaussian wavefunctional \cite{KSY90}, we describe the pairing
correlations by a suitable 
Bogoliubov unitary transformation. 
Furthermore, this could be the starting point for the further
approximation scheme beyond the Gaussian \cite{NTP83}.
In recent publications \cite{LTP92}, we have applied this scheme
to the $\phi^4$ theory and Chiral Gross-Neveu Model (hereafter
referred to as I,II). 

\indent
The main point of this paper is to apply this method within the
context of interacting fermion-boson models.
As a first step towards this end, Takano Natti and de Toledo Piza
\cite{TNTP97} have obtained relevant dynamics for the Jaynes-Cummings
Hamiltonian, a well known model in quantum optics \cite{JC63}. 
This can be seen as $0+1$ dimensional quantum field theory known as
relativistic scalar plasma \cite{Ka67}. 
Exact numerical results have been useful in this case to assess
the quality of various approximation schemes. 
In particular, we have shown that the method described here is a good
approximation whenever a moving Gaussian has enough freedom
to mimic the complicated evolution of the exact wave function.  
In this paper, we report an
aplication of the same technique to describe the real-time evolution for
a fermion field interacting with scalar boson in $3+1$ dimensions.
An outline of the paper is as follows.
In Sec. II we extend the discussions of I and II to a system countaing
interacting boson and fermion fields.  
Sec. III will illustre this scheme in the simplest context of scalar
plasma model.
In Sec. IV we discuss the self-consistent renormalization for the
resulting equations in the equilibrium situation.
The renormalization of the time-dependent equations is then considered
in Sec V and finite initial conditions discussed.

\medskip 
\renewcommand{\theequation}{\arabic{equation}}
\begin{center}
{\bf II. Kinetic of Interacting Fermion-Scalar Systems}
\end{center}
\smallskip

\indent{The} present discussion for the real-time evolution of theories
containing both scalar and fermion fields basically follows the
earlier works in the context of $\phi^4$ and Chiral-Gross-Neveu
models.   
The basic idea of our approach is to focus on the time evolution of the
expectation values of linear, $\phi(x)$, and bilinear field operators
such as $\phi(x)\phi(x)$ , $\bar\psi(x)\psi(x)$ , $\psi(x)\psi(x)$ and
etc, referred to as Gaussian observables.  
Many technical details of the method can be found in I and II, and we
restrict ourselves here to a few key steps of the derivation of
equations of motion for the present case.  

\indent{Let} us consider first the bosonic sector of the system. The
Heisenberg field operators $\phi(x)$ and $\Pi(x)$ are expanded as 
\begin{eqnarray}
\label{3}
\phi({\bf x},t) &=&\sum_{\bf p} \frac{1}{(2Vp_{0})^{1/2}} \left[b_{\bf
p}(t)e^{i{\bf p}.{\bf x}}+b_{\bf p}^{\dag}(t) e^{-i{\bf p}.{\bf
x}}\right]\;\;\nonumber\\ \\
\Pi({\bf x},t) &=&i\sum_{\bf p} \left(
\frac{Vp_{0}}{2}\right)^{1/2} \left[b_{\bf p}^{\dag}(t)e^{-i{\bf
p}.{\bf x}} -b_{\bf p}(t)e^{i{\bf p}.{\bf x}}\right]\;\;,\nonumber
\end{eqnarray}

\noindent
where $b_{\bf p}^{\dag}(t)$ and $b_{\bf p}(t)$ are the usual boson
creation and annihilation operators satisfying the standard
commutation relations at equal times
\begin{eqnarray}
\label{4}
[b_{\bf p}(t),b_{\bf p^{\prime}}^
{\dag}(t^{\prime})]_{t=t^{\prime}}= \delta_{{\bf p},{\bf
p^{\prime}}}\;\;, \;\;[b_{\bf p}^{\dag}(t),b_{\bf
p^{\prime}}^{\dag} (t^{\prime})]_{t=t^{\prime}}=[b_{\bf p}(t), b_{\bf
p^{\prime}}(t^{\prime})]_{t=t^{\prime}}=0\;\;.  
\end{eqnarray}

\noindent 
In Eq.(\ref{3}) $V$ is the volume of the periodicity box,

\[
(p_{0})^{2}={\bf p}^2+\Omega^{2}\;\;\;\; {\rm and}\;\;\;\;
px=p_{0}t-{\bf p}.{\bf x} \; ,
\]
\noindent
where $\Omega$ is the mass parameter and will be fixed
later. 

\indent{In} order to deal with condensate and pairing dynamics of the scalar
field we follow Sec. II of I and define a unitary Bogoliubov
transformation as
\begin{equation}
\label{4a}
\left[
\begin{array}{c}
\beta_{\bf p}\\
        \\
\beta_{-\bf p}^{\dag}
\end{array}
\right]=\left[
\begin{array}{cc}
\cosh \kappa_{\rm p}-i\frac{\eta_{\rm p}}{2}  &  
\sinh \kappa_{\rm p}-i\frac{\eta_{\rm p}}{2}\\ \\
\sinh \kappa_{\rm p}+i\frac{\eta_{\rm p}}{2} &  
\cosh \kappa_{\rm p}+i\frac{\eta_{\rm p}}{2}
\end{array}
\right]
\left[
\begin{array}{c}
d_{\bf p}  \\
      \\
d_{-\bf p}^{\dag}
\end{array}
\right]\;\;,
\end{equation}

\noindent
where $d_{\bf p}$ is the shifted boson operator
\begin{equation}
\label{4b}
d_{\bf p}(t)\equiv b_{\bf p}(t)- B(t)\delta_{{\bf
p},0}\;\;\;\;{\mbox{with}}\;\;\;\; 
B_{\bf p}(t) \equiv\langle b_{\bf p}(t) \rangle = Tr_{\mbox{\tiny
BF}}\;\left[b_{\bf p}(t){\cal F_0}\right]\;\;.
\end{equation}

\noindent{Here} and in what follows the symbol $Tr_{\mbox{\tiny BF}}$
denotes a trace over both bosonic and fermionic variables. Partial
traces over bosonic or fermionic variables will be written as 
$Tr_{\mbox{\tiny B}}$ and $Tr_{\mbox{\tiny F}}$ respectively. 
The Bogoliubov coefficients are determined from the secular problem
given in (2.8)-(2.10) of I to yield
$\langle\beta_{\bf p}\beta_{-\bf p}\rangle=0$. 
The resulting eigenvalues, 
$\nu_{\bf p}=\langle\beta^{\dag}_{\bf p}\beta_{\bf p}\rangle$, 
are quasi-particle occupation numbers.

\indent
For the fermion field, we follow the discussion of II and
decompose the field operators in Fourier components,
\begin{eqnarray}
\label{5}
\psi({\bf x},t)&=&\sum_{{\bf
k},s}\left(\frac{M}{k_{0}}\right)^{1/2}\frac{1}{\sqrt V}
\left[u_{1}({\bf k},s)a_{{\bf k},s}^{(1)}(t)e^{i {\bf k}.{\bf x}}
+u_{2}({\bf k},s){a_{{\bf k},s}^{(2)}}^{\dag}(t) e^{-i {\bf k}.{\bf
x}}\right]\;\; \nonumber\\ \\ \bar\psi({\bf
x},t)&=&\sum_{{\bf
k},s}\left(\frac{M}{k_{0}}\right)^{1/2}\frac{1}{\sqrt V}\left[\bar
u_{1}({\bf k},s){a_{{\bf k},s}^{(1)}}^{\dag}(t)e^{-i {\bf k}.{\bf x}}+
\bar u_{2}({\bf k},s)a_{{\bf k},s}^{(2)}(t) e^{i {\bf k}.{\bf
x}}\right]\;\;, \nonumber
\end{eqnarray}

\noindent where  
${a_{{\bf k},s}^{(1)}}^{\dag}(t)$ and $a_{{\bf k},s}^{(1)}(t)$
[${a_{{\bf k},s}^{(2)}}^{\dag}(t)$ and $a_{{\bf k},s}^{(2)}(t)$] are
fermion creation and annihilation operators associated with positive
[negative]-energy solutions $u_{1}({\bf k},s)$ [$u_{2}({\bf k},s)$]
and satisfy the anticommutation rules
\begin{eqnarray}
\label{6}
\{{a_{{\bf k},s}^{(\lambda)}}^{\dag}(t),a_{{\bf
k^{\prime}},s^{\prime}}^{(\lambda^{\prime})}
(t^{\prime})\}_{t=t^{\prime}}&=&\delta_{{\bf k},{\bf
k^{\prime}}}\delta_{s,s^{\prime}}\delta_{\lambda,\lambda^{\prime}}
\;\;\;\;{\rm for}\;\;\;\lambda,\lambda'=1,2\;\; \nonumber\\ \\
\{{a_{{\bf k},s}^{(\lambda)}}^{\dag}(t),{a_{{\bf
k^{\prime}},s^{\prime}}^{(\lambda^{\prime})}}^{\dag}
(t^{\prime})\}_{t=t^{\prime}}&=&\{a_{{\bf k},s}^{(\lambda)}(t),
a_{{\bf k^{\prime}},s^{\prime}}^{(\lambda^{\prime})}
(t^{\prime})\}_{t=t^{\prime}}=0\;\;.  \nonumber
\end{eqnarray}
\vskip 0.2cm

\noindent 
In Eq.(\ref{5}) the spinors are normalized to $k_{0}/M$ and
\be
\label{6a}
(k_{0})^{2}={\bf k}^{2}+M^{2}\;\;\;\; {\rm and}\;\;\;\; kx=k_{0}t-{\bf
k}.{\bf x} \;
\ee

\noindent
where $M$ is a mass parameter for the fermions. 

\indent
Adapting Eqs. (3)-(11) and (30) of II to the present case of
3+1 dimensional field we can write the  Bogoliubov
quasi-particle operators as 
\begin{equation}
\label{5a}
\left[\!
\begin{array}{c}
      \alpha_{{\bf k},s}^{(1)}    \\
                   \\
      \alpha_{{\bf k},s}^{(2)}    \\
                   \\
{\alpha_{-{\bf k},{\bar s}}^{(1)}}^{\dag}\\
                   \\
{\alpha_{-{\bf k},{\bar s}}^{(2)}}^{\dag}
\end{array}
\!\right]\!=\!\left[\!
\begin{array}{cccc}
\cos\varphi_{\bf{k}} &      0       &      0       & 
          e^{-i\gamma_{\bf{k}}}\sin\varphi_{\bf{k}} \\
          &              &              &              \\  
  0       & \cos\varphi_{\bf{k}} & -e^{-i\gamma_{\bf{k}}}
\sin\varphi_{\bf{k}} &     0        \\ 
          &              &              &              \\ 
  0       & e^{i\gamma_{\bf{k}}}\sin\varphi_{\bf{k}}
          &\cos\varphi_{\bf{k}} &     0        \\ 
          &              &              &              \\
-e^{i\gamma_{\bf{k}}}\sin\varphi_{\bf{k}} &      0       &      0
          & \cos\varphi_{\bf{k}} 
\end{array}
\!\right]\left[
\!\begin{array}{c}
    a_{{\bf k},s}^{(1)}    \\
                         \\ 
    a_{{\bf k},s}^{(2)}    \\
                         \\
{a_{-{\bf k},{\bar s}}^{(1)}}^{\dag}\\
                         \\
{a_{-{\bf k},{\bar s}}^{(2)}}^{\dag}
\end{array}
\!\right]
\end{equation}

\noindent
and require 
$\langle \alpha_{-{\bf k},\bar s}^{(\lambda^{\prime})}
\alpha_{{\bf k},s}^{(\lambda)}\rangle = 0$
in order to determine this unitary transformation. 
The quasi-fermion occupation number will be given by
$\nu^{(\lambda)}_{{\bf k},s} 
=\langle{\alpha^{(\lambda)}_{{\bf
k},s}}^{\dag}\alpha^{(\lambda)}_{{\bf k},s}\rangle$.  
In (\ref{5a}) we used 
$|\bar{s}\rangle=(-)^{1/2-s}|-s\rangle$ to write the spin labels. 

\indent
The next step is to obtain the equations of motion for the condensate,
$\langle\phi\rangle$ and $\langle\Pi\rangle$,
and for the Bogoliubov parameters given in (\ref{4a})-(\ref{4b}) and
(\ref{5a}). 
This proceeds by using the Heisenberg equation of motion (\ref{1a}), where
$\cal O$ are performed by linear or bilinear combinations of creation
and annihilation parts of the field operators
[see (2.17) of I and (5) of II for details].
For the bosonic variables we have
\begin{eqnarray}
\label{10}
i \langle\dot\phi\rangle&=&Tr_{\mbox{\tiny BF}}[\phi,H]{\cal
  F}_{0}\\\nonumber\\ 
\label{10a}
i \langle\dot\Pi\rangle&=&Tr_{\mbox{\tiny BF}}[\Pi,H]{\cal
  F}_{0} \\\nonumber\\
\label{8}
\dot \nu_{\rm p} &=& Tr_{\mbox{\tiny BF}}[\beta_{\bf
p}^{\dag}\beta_{\bf p},H] {\cal F}_{0}\\ \nonumber\\ 
\label{9}
-i\dot\kappa_{\rm p}
-e^{-\kappa_{\rm p}}\left(\dot\eta_{\rm p}+\dot\kappa_{\rm p}\eta_{\rm
    p}\right) 
&=& \frac{Tr[\beta_{\bf p}^{\dag}\beta_{-\bf p}^{\dag},H] {\cal F}_{0}}
{1+2\nu_{\rm p}} \;\;.
\end{eqnarray}

\noindent
Analogously, the dynamical equations for
fermionic variables read as
\begin{eqnarray}
\label{11}
\dot\nu^{(\lambda)}_{{\bf k,},s}
&=&Tr_{\mbox{\tiny BF}}
\left([{\alpha_{{\bf k},s}^{(\lambda)}}^{\dag}\alpha_{{\bf k},s}^{(\lambda)},
H]{\cal F}_{0}\right) \\
\nonumber\\
\label{12}
\left[i\dot\varphi_{\bf k}+\dot\gamma_{\bf k}\sin\varphi_{\bf k}
\cos\varphi_{\bf k}\right]e^{-i\gamma_{\bf k}}
&=&
\frac
{Tr_{\mbox{\tiny BF}}\left([\alpha_{-{\bf k},s}^{(1)}\alpha_{{\bf k},s}^{(2)},
H]{\cal F}_{0}\right)}{1-\nu_{{\rm k},s}^{(1)}-\nu_{{\rm k},s}^{(2)}}
\;\;.
\end{eqnarray}

\indent
Our implementation of the Gaussian mean-field approximation consists
in assuming a factorized form for the density matrix \cite{TNTP97}
\be
\label{32c}
{\cal F}_{0}(t)
={\cal F}^{\mbox{\tiny B}}_{0}{\cal F}^{\mbox{\tiny F}}_{0}\;\;.
\ee

\noindent{In} Eq.(\ref{32c}) the subsystem densities 
${\cal F}^{\mbox{\tiny B}}_{0}$ and ${\cal F}^{\mbox{\tiny F}}_{0}$ 
are gaussian densities given in Eq.(\ref{1b}), now rewritten in the
diagonal forms using the Bogoliubov quasiboson and quasifermion
operators,
\begin{equation}
\label{32a}
{\cal F}^{\mbox{\tiny B}}_{0}=\prod_{\bf p}\frac {1} {1+\nu_{\bf
p}}\left(\frac {\nu_{\bf p}}{1+\nu_{\bf p}}\right) ^{\beta_{\bf
p}^{\dag}\beta_{\bf p}}\;\;,
\end{equation}
\vskip 0.2cm
\begin{equation}
\label{32b}
{\cal F}^{\mbox{\tiny F}}_{0}=\prod_{{\bf k},s,\lambda} [\nu_{{\bf
k},s}^{(\lambda)}{\alpha^{(\lambda)}_{{\bf k},s}}^{\dag}
\alpha^{(\lambda)}_{{\bf k},s}+(1-\nu_{{\bf k},s}^{(\lambda)})
\alpha_{{\bf k},s}^{(\lambda)}{\alpha^{(\lambda)}_{{\bf
k},s}}^{\dag}]\;\;.
\end{equation}
 
\noindent
Substituting ${\cal F}_{0}$ into (\ref{8}) and (\ref{11}) one verifies
immediately that the occupation numbers are constant,
\be
\dot \nu_{\rm p} 
= Tr_{\mbox{\tiny BF}} [\beta_{\bf p}^{\dag}\beta_{\bf p},H] {\cal
  F}_{0} 
=\dot\nu_{{\bf k},\lambda}
=Tr_{\mbox{\tiny BF}}
[{\alpha_{{\bf k},s}^{(\lambda)}}^{\dag}\alpha_{{\bf k},s}^{(\lambda)},
H]{\cal F}_{0} =0 \;\;.
\ee

\noindent
Thus, Eqs.(\ref{10})-(\ref{12}) form a set of self-consistent equations of 
motion to the Gaussian variables. 
Its implementation for a specific model consists essentially in taking
the traces of appropriate commutators in the Fock space.
The fact that occupancies are constant implies that
the entropy function associated with ${\cal F}_{0}$, i.e.,
$S=-Tr_{\mbox{\tiny BF}}{\cal F}_{0}\log {\cal F}_{0}$, does not
change in time.
One recovers therefore the usual isoentropic character of the Gaussian
approximation.  

\indent
A method to improve the mean-field approximation in field theory
was discussed in I. 
The approach follows the line of thinking of a time-dependent
projection technique proposed some time ago by Willis and Picard
\cite{WP74} in the context of master equation for coupled systems 
and extended later by Nemes and de Toledo Piza to study
nonrelativistic many-fermion dynamics \cite{NTP83}. 
The method consists essentially in writting the correlation
information of the {\it full density} of the system in terms of a
memory kernel acting on the uncorrelated density ${\cal} F_{0}$, with
the help of a time-dependent projector (see Sec. III and Appendix A of
I for details). 
At this point, an systematic mean-field expansion for two-point
correlations can be perfomed. 
The lowest order reported in this work corresponds to the
results of the usual Gaussian approximation. 
The higher orders describe the dynamical correlation effects between
the subsystems and are expressed through suitable memorial integrals
added to the mean-field dynamical equations. 
Thus, the resulting equations acquire the structure of
kinetic equations, with the memory integrals performing as {\it
collisional} dynamics terms which eliminate the isoentropic
constraint. This higher order approximation has been implemented for
the case of $0+1$ dimensional field model \cite{TNTP97} and will not
be done in this paper.

\begin{center}
{\bf III. Effective Dynamics of Relativistic Scalar Plasma}
\end{center}

\indent 
The procedure discussed in the preceding section is quite general and
can be, in principle, utilized for systems containing scalar and
spin-1/2 particles. 
This is not, however a straightforward task, since divergences appear in
the equations require a discussion of the renormalizability of
specific models of interest.  
In Ref.\cite{SHR86} properties and stability of possible vacuum states
for  several fermion-scalar models utilizing the
Gaussian-effective-potential (GEP) approach have been studied.   
It is shown that in 3+1 dimensions the models seem doomed to
instability and the GEP is bounded below only if the Yukawa coupling
vanishes.  
We have reached a similar conclusion in an early stage of this work.

\indent
In order to give a well defined physical application, we shall adopt
here the relativistic scalar plasma  model widely used within the 
context of ultrahigh dense matter \cite{AH84}.  
This system consists of relativistic fermion gas interacting through a
massive scalar meson only and corresponds to one of the simplest
quantum-field theoretical models used to discuss the relativistic
dense matter in the contexts of heavy-ion collisions and the
high-density astrophysical system \cite{Ka67}. 
In this application, one considers the mean-field
contributions, $\langle\phi\rangle$, are much larger than the flutuactions,
$\langle\phi^2\rangle - \langle\phi\rangle ^2$. Therefore, the bosonic
pairing is ignored and Bogoliubov transformation (\ref{4a}) reduces to
identity. 
We shall show in the next sections that the resulting equations of
interest in this case can be made finite by introducing appropriate
counterterms. 
 
\indent
We now proceed to apply the technique discussed in Sec. II to the 
scalar plasma model. 
The dynamics is governed by the Hamiltonian (we use the notation:
$\int_{\bf x}\equiv\int d^{3}x$)
\begin{eqnarray}
\label{1}
H&=&\int_{\bf x} {\cal H}\;,\nonumber\\ \\ {\cal H}&=&-\bar\psi
(i\vec\gamma.{\vec\partial}-m)\psi-
g\bar\psi\phi\psi+\frac{1}{8\pi}
\left[ (4\pi \Pi)^2 + |\partial\phi|^2+
\mu^2\phi^2\right]+{\cal H}_{c}\;\;, \nonumber
\end{eqnarray}
\vskip 0.5cm
\noindent 
where the parameters $m$ and $\mu$ are, respectively, the mass of
fermion and scalar particles and $g$ is the coupling constant. The
last term of this expression contains the counterterms necessary to remove
the infinities occurring later \cite{AH84}, 

\begin{equation}
\label {27}
4\pi {\cal H}_{\mbox {\tiny C}} = \frac{A}{1!}\phi + 
\frac{\delta \mu^{2}}
{2!}\phi^{2} + \frac{C}{3!}\phi^{3} + \frac{D}{4!}\phi^{4} \;\;,
\end{equation}
\vskip 0.5cm
\noindent
where the coeffcients $A$, $\delta \mu^{2}$, $C$ and $D$ are
infinite constants to be defined later. 

\indent
Introducing these ingredients into Eqs.(\ref{10}), (\ref{10a}) and
(\ref{12}) 
one easily obtains dynamical equation for the condensate
\begin{eqnarray}
\label{63}
\langle\dot\phi\rangle &\!\!=&\!\!4\pi\langle\Pi\rangle\\\nonumber\\
\label{121}
\langle\dot\Pi\rangle
&=&-\frac{1}{4\pi}\left[A+\frac{C}{2}G(\Omega)\right]
   -\frac{1}{4\pi}\left[\mu^2+\delta\mu^2+\frac{D}{2}G(\Omega)\right]
\langle\phi\rangle
-\frac{C}{8\pi}\langle\phi\rangle^2-\frac{D}{24\pi}\langle\phi\rangle^3
\nonumber\\\nonumber\\
&\!\!-&\!\!g
\sum_{s} \int_{\rm k} \frac{1}{k_{0}}\left[M\cos 2\varphi_{{\rm k}}+
k\sin 2\varphi_{{\rm k}}\cos\gamma_{{\rm
k}}\right] (1-\nu_{ {\rm k},s}^{(1)}-\nu_{{\rm
k},s}^{(2)})\;\;,
\end{eqnarray}

\vspace{0.3cm}
\noindent
where $k\equiv |{\bf k}|$, and two real equations for the fermion
pairing 
\begin{eqnarray}
\label{73}
\dot \varphi_{{\rm k}}&=& \frac{k}{k_{0}}(M-\bar{m})
\sin\gamma_{{\rm k}}\\\nonumber\\
\label{74}
\sin2\varphi_{{\rm k}}\dot\gamma_{{\rm k}}&=&
\frac{2({\bf k}^2+\bar{m}M)}{k_{0}}\sin2\varphi_{{\rm k}} 
+\frac{2(M-\bar{m})}{k_{0}}k 
\cos2\varphi_{{\rm k}}\cos\gamma_{{\rm k}}\;\;.
\end{eqnarray}

\noindent 
In (\ref{121}) we have introduced the notation
\begin{equation}
\label{87a}
G(\Omega) =
\int_{\bf p} \frac{1+2\nu_{\rm p}}{2\sqrt{{\bf p}^2+{\Omega}^2}}
=\frac{1}{8\pi^2}\left(\Lambda^2_{\bf p} 
+{\Omega}^2\log\frac{2\Lambda_{\bf p}}{\sqrt{e}{\Omega}}\right) 
+\int_{\bf p} \frac{\nu_{\rm p}}{p_{0}}  \;\;
\end{equation}

\noindent
where $\Lambda_{\bf p}$ is a momentum cutoff used in evaluating the
integral. In (\ref{73})-(\ref{74}) 
$\bar m\equiv m-g\langle\phi\rangle$
stands for the effective fermion mass.

\indent
Another physical quantity of interest is the energy density of the system, 
\begin{eqnarray}
\label{123a}
\frac{\langle H\rangle}{V} &=& \frac{1}{V}Tr H{\cal F}_{0}
\nonumber\\\nonumber\\
&=&-\sum_{s}\int_{\rm k} \left[\frac{({\bf k}^2+\bar m{M})}{k_{0}} \cos
2\varphi_{{\rm k}}+\frac{(\bar m-M)}{k_{0}}k \sin 2\varphi_{\rm
k}\cos\gamma_{\rm k}\right](1-\nu_{\rm k}^{(1)}- \nu_{\rm
k}^{(2)})\nonumber\\\nonumber\\
&&+\frac{1}{8\pi}\left[\langle\Pi\rangle^2+
\mu^2\langle\phi\rangle^2\right]
+\frac{1}{4\pi}\left[A+\frac{C}{2}G(\Omega)\right]
\langle\phi\rangle
+\frac{1}{8\pi}\left[\delta\mu^2+\frac{D}{2}G(\Omega)\right]\langle\phi\rangle^2
\nonumber\\\nonumber\\
&&+\frac{C}{24\pi}\langle\phi\rangle^3
+\frac{D}{96\pi}\langle\phi\rangle^4
+\frac{1}{8\pi}\left[\mu^2+\delta\mu^2\right]G(\Omega)
+\frac{D}{32\pi}G^2(\Omega)\;\;.
\end{eqnarray}
 
\noindent
An important feature of this scheme is that the mean energy is
conserved, a property which can be verified explicitly using the equations of
motion (\ref{63})-(\ref{74}). 
Notice also that the results above contain divergent integrals,
so that the coefficients $A$, $\delta \mu^{2}$, $C$ and $D$ have to be
adjusted appropriately in order to give a well defined dynamics. 
We shall focus on this point in the next sections.

\smallskip
\begin{center}
{\bf IV--Static Equations and Renormalization}
\end{center}
\smallskip

This section considers Eqs. (\ref{63})-(\ref{74}) in the
equilibrium situation. 
We shall investigate the solution of these equations and study
renormalization conditions. 
Hence, we set 
\begin{equation} 
\dot\gamma_{\rm k}=\dot\varphi_{\rm k}
=\langle\dot\phi\rangle= \langle\dot\Pi\rangle=0\;\;.
\end{equation}

\noindent
Since our main goal here is to set up a Gaussian mean field
approximation for interacting fermion-boson fields, we shall, for
simplicity, not consider the matter contribution in the
present calculation.
This can be done in the case of Gaussian mean-field approximation by
assuming an appropriate statistical distribution to the occupation
numbers $\nu_{\rm  p}$ and $\nu_{\rm k,s}^{(\lambda)}$ as initial
conditions in (\ref{8}) and (\ref{11}). 
Thus, for the rest of discussion we have vanishing occupancies, 
$\nu_{\rm  p}=\nu_{\rm  k,s}^{(\lambda)}=0$, and the sum in spin gives a
numerical factor, $\sum_{s}=2$. 
With these assumptions, Eqs.(\ref{63})-(\ref{74}) become

\begin{eqnarray}
\label{81}
&&\langle\Pi\rangle|_{\mbox{\tiny eq}}=0\;,\\\nonumber\\ 
\label{82}
&&\left[A+\frac{C}{2}G(\Omega)\right]+
\left[\mu^2+\delta\mu^2+\frac{D}{2}G(\Omega)\right]
\langle\phi\rangle|_{\mbox{\tiny eq}}
+\frac{C}{2}\langle\phi\rangle|_{\mbox{\tiny eq}}^2
+\frac{D}{6}\langle\phi\rangle|_{\mbox{\tiny eq}}^3\nonumber\\\nonumber\\
&&+8\pi g\int_{\rm k}\frac{1}{k_{0}}\left( M\cos 2\varphi_{\rm
k}|_{\mbox{\tiny eq}}+k \sin 2\varphi_{\rm
k}|_{\mbox{\tiny eq}}\cos\gamma_{\rm k}|_{\mbox{\tiny eq}}
\right) = 0 \;, \\\nonumber\\
\label{77}
&&(M-\bar m) \sin\gamma_{\rm k}|_{\mbox{\tiny eq}}=0 \;,\\\nonumber\\ 
\label{78}
&&({\bf k}^{2}+\bar{m}M)\sin 2\varphi_{\rm k}|_{\mbox{\tiny eq}}
+(M-\bar m)\cos 2\varphi_{\rm k}|_{\mbox{\tiny eq}}
\cos\gamma_{\rm k}|_{\mbox{\tiny eq}} = 0 \;,
\end{eqnarray}

\noindent
where $\bar{m}$ is now the effective fermion mass calculated at
$\langle\phi\rangle|_{\mbox{\tiny eq}}$. 
Solutions of this set of equations correspond to a stationary points
of the energy density, and include the ground state of the system in
this approximation. 

\smallskip 
\centerline{\bf A. Equilibrium conditions and mass parameters}
\smallskip

\indent
Considering equations (\ref{77}) and (\ref{78}) one sees that two
situations have to be analysed.\\
\indent i)$\;\;\sin\gamma_{\rm k}|_{\mbox{\tiny eq}}=0\;\;$ \\
\indent Using this solution in Eq.(\ref{78}) we have  
\be
\label{82d}
\cos2\varphi_{\rm k}|_{\mbox{\tiny eq}}
=\pm\frac{({\bf k}^2+\bar mM)}{k_{0} \bar k_{0}}\;\;,\;\;\;\;
\sin 2\varphi_{\rm k}|_{\mbox{\tiny eq}}
=\mp\frac{M-\bar m} {k_{0} \bar k_{0}}
\ee

\noindent where $\bar k_{0}=\sqrt{{\bf k}^2+\bar m^2}$. Substituting this
result into (\ref{82}) one finds for this case the following gap
equation 
\begin{eqnarray}
\label{89}
\left[A+\frac{C}{2}G(\Omega)\right]
+\left[\mu^2+\delta\mu^2+\frac{D}{2}G(\Omega)\right]
\langle\phi\rangle|_{\mbox{\tiny eq}}
+\frac{C}{2}\langle\phi\rangle|_{\mbox{\tiny eq}}^2+\frac{D}{6}
\langle\phi\rangle|_{\mbox{\tiny eq}}^3= \pm16\pi g \bar m G({\bar m})\;,
\end{eqnarray}

\noindent 
where $G({\bar m})$ is the divergent integral defined in (\ref{87a})
with $\nu_{\rm p}=0$.

\indent ii)$\;\;M=\bar m$\\
\indent In this case $\gamma_{\rm k}$ can have any value and
(\ref{78}) has as solution
\begin{equation} 
\label{82a}
\sin2\varphi_{\rm k}|_{\mbox{\tiny eq}}=0\;\;,\;\;\;
\cos2\varphi_{\rm k}|_{\mbox{\tiny eq}}=\pm1\;\;\; 
\end{equation}

\noindent
Comparing now (\ref{82d}) with
(\ref{82a}) one notices immediately that the former includes the
second solution as a particular case, where the mass parameter
$M$ is defined here by the equilibrium condition and Bogoliubov 
transformation (\ref{5a}) reduces to an identity matrix.  

\indent
The above discussion shows that the equilibrium value
$\langle\phi\rangle|_{\mbox{\tiny eq}}$ of the condensate, or the
effective fermion mass $\bar m$ are independent of the initial choice
of the mass paramete $M$. 
In fact, the effects of pairing transformation will be such that the
effective fermion mass is always given by the equilibrium value
calculated with (\ref{89}).   
(See also (\ref{110}) below for the  mean-field energy).
Henceforth, we shall thus take $M=m$ without loss of generality. 

\medskip
\centerline{\bf B. Counterterms}
\medskip

\indent
The arbitrary parameters $A$, $\delta\mu$, $C$ and $D$ can be
fixed easily by adjusting the coefficients of
$\langle\phi\rangle|_{\mbox{\tiny eq}}$ in both of sides of
(\ref{89}) to give a finite gap equation. Thus, taking account the
$\pm$ signs in (\ref{89}) we choose the
following self-consistent renormalization prescription
\begin{eqnarray}
\label{94}
D&=&\pm48\pi g^4 L(m)\;,\\
\label{95}
\delta\mu^2&=&\mp24\pi^2 g^4L(m)G(\mu)\mp16\pi g^2G(0)\pm24\pi m^2g^2 L(m)\;,\\
\label{96}
C&=&\mp48\pi mg^3L(m)\;,\\
\label{97}
A&=&\pm24\pi mg^3L(m)G(\mu)\pm16\pi mgG(m)\;,
\end{eqnarray}

\noindent
where

\begin{equation}
\label{98}
L(m) \equiv \int_{\bf k} \frac{1}{2{\bf k}^2 ({\bf k}^2+m^2)^{1/2}}
=\frac{1}{4\pi^2}\log\frac{2\Lambda_{\bf p}}{m}\;.
\end{equation}
\vskip 0.3cm
\noindent 
Furthermore, $\mu$ and $m$ are the mass scales for boson and
fermion fields respectively. 

\indent Substitution of these counterterms into (\ref{89}) produce the
appropriate cancelations which render the equation finite. 
A combination of type $ L(m)[G(\mu)-G(\Omega)] $ comes from the first
two terms.  
Since $\Omega$ is an arbitrary expansion mass parameter, one can
remove this divergence by setting $\Omega=\mu$. 
The resulting finite gap equation is

\begin{equation}
\label{103}
\frac{\pi}{2}{\mu}^2\langle\phi\rangle|_{\mbox{\tiny eq}}- g{\bar
m}^3\left[\ln\left(\frac{\bar m}{m}\right)+\frac{1}{2}\right]=0\;\;.
\end{equation}
\vskip 0.3cm

\noindent The above equation together with (\ref{77})-(\ref{78}) 
determine the stationary points of the model in the
Gaussian mean-field approximation.

\medskip 
\centerline{\bf C. Mean energy and stationary solutions}  
\medskip

We examine next the energy density when it is stationary
with respect to the fermion variables. 
With the help of (\ref{82d}), Eq.(\ref{123a}) becomes 

\begin{eqnarray}
\label{110}
\frac{\langle H\rangle}{V}
\bigl( \varphi_{\rm k}|_{\mbox{\tiny eq}}, \gamma_{\rm
  k}|_{\mbox{\tiny eq}},\langle\phi\rangle\bigr)
  &\!\!=&\!\!
\mp\int_{\rm k}G^{-1}(\bar m)+
\frac{1}{4\pi}\left[A+\frac{C}{2}G(\mu)\right] 
\langle\phi\rangle
+\frac{1}{8\pi}\left[\mu^2+\delta\mu^2+
\frac{D}{2}G(\mu)\right]
\langle\phi\rangle^2\nonumber\\\nonumber\\
&\!\!+&\!\!\frac{C}{24\pi}\langle\phi\rangle^3+
\frac{D}{96\pi}\langle\phi\rangle^4 \;\;.
\end{eqnarray}

\noindent
Substituting next the counterterms (\ref{94})-(\ref{97})
into this result, we have the renormalized
version of mean-field energy as function of mean-field value
$\langle\phi\rangle$ or $\bar{m}$ 

\begin{equation}
\label{112}
\frac{\langle H\rangle}{V}=\frac{1}{8\pi^2}\left[\frac{\pi\mu^2}{g^2}
(m-{\bar m})^2+{\bar m}^4\ln\left(\frac{\bar m}{m}\right)
+\frac{({\bar m}^4-m^4)}{4}\right]\;\;,
\end{equation}
\vskip 0.3cm

\noindent where we have added appropriate constant in order to get
$\langle H\rangle(\langle\phi\rangle=0)=0$. 
One can now discuss the possible solution of (\ref{103}) by analysing
the minima of (\ref{112}). 

\indent Let us define $x\equiv g\langle\phi\rangle|_{\mbox{\tiny
eq}}/m$ and $E(x)\equiv (8\pi^2/m^4)\langle H\rangle/V$. Eqs. (\ref{103}) and
(\ref{112}) can then be written respectively as

\begin{equation}
\label{115}
\frac{\pi\mu^2}{2g^2m^2}x-(1-x)^3\left[\ln(1-x)+\frac{1}{2}\right]=0\;\;
\end{equation}

\noindent and

\begin{equation}
\label{113}
E(x)=\frac{\pi\mu^2}{g^2m^2}x^2 +(1-x)^4
\left[\ln(1-x)+\frac{1}{4}\right]-\frac{1}{4}\;\;.
\end{equation}
\vskip 0.3cm

\noindent
The combination $\frac{4g^2m^2}{\pi\mu^2}$ has been used as an effective
coupling constant in Ref.\cite{AH84}.   

\indent 
The behavior of $E(x)$ is shown in Figs.(1)-(4) for several
combination  of $\mu/m$ and $g^2$.  
Notice first that this function has the domain at $0\le x \le 1$,
which is the physical range for the fermion mass. 
The point $x=0$ corresponds to $\langle\phi\rangle=0$ or $\bar m=m$
and $x=1$ is the case when the effective fermion mass $\bar m=0$. 
Qualitatively, the results indicate that the system always presents a
single {\it minimum}.  
Figs.(1)-(2) show $E(x)$ for several values of $g^2$ with $\mu/m$
fixed (see figure captions for numerical values of the parameters). 
The positions of $x_{min}$ indicate that the vacuum approaches
$x=0$ when we decrease $g^2$.  
In the limit of $g^2\rightarrow 0$, one gets $x_{min}\rightarrow 0$,
as shown in Fig.(3). 
In this case, $m\approx\bar{m}$ is the optimal fermion mass, as it
must be in the free field theory. 
On the other hand, Figs.(3)-(4) plot the function $E(x)$ keeping
the value of the Yukawa coupling $g^2$ fixed, but with different
values of the ratio $\mu/m$. 
Comparing these two curves one sees $x_{min}\rightarrow0$ when
$\mu/m\rightarrow\infty$. 
In other words, when the meson mass is large, the force range is
small, as usual in the Yukawa theory. 
In the limit of infinity $\mu$, the fermion particles of the system
cannot interact. 
It can be seen also from the AH's formula, where $m/\mu$ plays roles
of effective coupling constant. 
The above discussions suggest that the field has always a stable
vacuum. 
This means that there is a finite range around the minimum where the
dynamics of the system is well defined. 
In the next section, we shall therefore discuss renormalization conditions
for the time-dependent equations. 

\smallskip
\begin{center}
{\bf V. Renormalization and Initial Conditions for
            the Time-Dependent Equations}
\end{center}
\smallskip

The last section has discussed the problem of renormalization
for the vacuum sector of the relativistic scalar plasma model in the
Gaussian mean-field approximation. 
We have shown that the physical quantities can be made finite with the
counterterms introduced in (\ref{27}) and (\ref{94})-(\ref{97}). 
Here we shall consider an off-equilibrium situation and
study renormalizability for the Gaussian equations of motion for this
model.   
(See, e.g., Refs.\cite{PS87,KL95} for the issue of renormalization of
time-dependent equations in $\phi^4$ field theory) 

\medskip
\centerline{\bf A.  Equations of Motion and Initial State}
\medskip

\indent
In order to pose the problem more clearly, let us begin by rewriting
Eq.(\ref{121}), with the help of (\ref{94})-(\ref{97}), as 
\begin{eqnarray}
\label{61a}
\langle\ddot\phi\rangle
=-\mu^2\langle\phi\rangle-8\pi g {\cal F}(t) \;\;.
\end{eqnarray}

\noindent
where
\begin{eqnarray}
\label{62a}
{\cal F}(t)=\int_{\rm k}{\cal F}_{\rm k}(t)
=\int_{\rm k}\left[\frac{m}{k_{0}}\cos 2\varphi_{{\rm k}}
+\frac{k}{k_{0}}\sin 2\varphi_{{\rm k}}\cos\gamma_{{\rm
k}}+\frac{\bar m(t)}{k}-\frac{\bar m^3(t)}{2k^2k_{0}}\right]\;\;.
\end{eqnarray}

\noindent
Notice that the integral is divergent unless ${\cal F}_{\rm k}$
decreases faster than $k^{-3}$. For simplicity, we shall ignore here
the possibilities of fractional powers, logarithmic or oscillatory
behavior and take the convergence condition for the integral to be  
\be
\label{63f}
\lim_{k \rightarrow \infty} {\cal F}_{\rm k} 
={\cal O}\left( \frac{1}{k^{4}} \right)\;\;.
\ee

\noindent
Thus, the allowed domains for the dynamical variables in momentum
space suffer strong constraints, and our investigation of
renormalizability of the 
equations of motion consists in analysing the large momentum behavior of 
${\cal F}_{\rm k}$ from this point of view.
In order to satisfy (\ref{63f}) at all times the pairing variables
$\varphi_{\rm k}$ and $\gamma_{\rm k}$ must have their time evolution
restricted. These restrictions are analysed better by making the variable
changes   
\be
\label{626}
k^2_0\cos 2\varphi_{{\rm k}}
=-mR_{\rm k}-k\sqrt{k^2_0-R^2_{\rm k}}\\\nonumber\\
\ee

\noindent or, equivalently,
\be
\label{627}
k^2_0\sin 2\varphi_{{\rm k}}
=-kR_{\rm k}+m\sqrt{k^2_0-R^2_{\rm k}}\;\;,
\ee

\noindent and
\be
\label{628}
\cos\gamma_{\rm k} = 1-W_{\rm k} \;\;.  
\ee

\noindent In terms of these new variables we have 
\be
\label{628a}
{\cal F}(t)
\;=\;\int_{\rm k}{\cal F}^{(1)}_{\rm k}+\int_{\rm k}{\cal F}^{(2)}_{\rm k}\;\;,
\ee

\noindent where
\begin{eqnarray}
\label{628b}
{\cal F}^{(1)}_{\rm k}&=&-\frac{R_{\rm k}}{k_0}
+\frac{\bar m(t)}{k}-\frac{\bar m^3(t)}{2k^2k_{0}} \;\;,\\\nonumber\\
\label{628c}
{\cal F}^{(2)}_{\rm k}&=&\frac{k}{k^3_0}W_{\rm k}
\left(-kR_{\rm k}+m\sqrt{k^2_0-R^2_{\rm k}}\right) \;\;.
\end{eqnarray}

\noindent
We see in ${\cal F}^{(1)}_{\rm k}$ that the leading term of 
$R_{\rm k\rightarrow \infty }$ must be $\bar m$ in order to cancel the
quadratic divergence. Thus, we define
\be
\label{629}
\lim_{k \rightarrow \infty}R_{\rm k} = \bar m + S_{\rm k} \;
\ee

\noindent
valid for large $k$. Using this in (\ref{628b}) one gets 
\bea
\label{630a}
\lim_{k \rightarrow \infty} {\cal F}^{(1)}_{\rm k}
&=&\frac{\bar m m^2}{k k_{0} ( k+k_{0} )}
-\frac{\bar m^3}{2k^2k_0}
-\frac{S_{\rm k} }{k_{0}}\nonumber\\
&=&\frac{1}{k_0}
\Bigl[\frac{\bar m(m^2-\bar m^2)}{ k ( k+k_0 ) }-S_{\rm k}\Bigr]
+ {\cal O}\left( \frac{1}{k^5} \right) \;\;.
\eea

\noindent
There is still a logarithmic divergence left. To keep this
under control  we must have
\be
\label{631}
\lim_{k \rightarrow \infty} S_{\rm k}
= \frac{\bar m(m^2-\bar m^2)}{ k ( k+k_0 ) }
\ee 

\noindent
in order to get finite $\int_{\rm k}{\cal F}^{(1)}_{\rm k}$.

\indent 
We examine next the convergence condition for $\int_{\rm k}{\cal
F}^{(2)}_{\rm k}$. 
Using (\ref{629}) and (\ref{631}) in (\ref{628c}) we find the
asymptotic expression for ${\cal F}^{(2)}_{\rm k}$
\be
\label{632}
\lim_{k \rightarrow \infty} {\cal F}^{(2)}_{\rm k}
\;=\;\left[\frac{k^3(\bar{m}^2-m^2)}{k^3_0(k\bar{m}+k_0m)}
+{\cal O}\left( \frac{1}{k^3} \right)
\right] W_{\rm k} \;\;.
\ee

\noindent
This will constrain the large-$k$ behavior of $W_{\rm k}$ to be
\be
\label{633}
\lim_{k \rightarrow \infty} W_{\rm k}
={\cal O}\left( \frac{1}{k^{3}} \right) \;\;,
\ee

\noindent
in order to satisfy (\ref{63f}).

\indent
The asymptotic conditions (\ref{629}), (\ref{631}) and (\ref{633})
restrict the dynamics of the Gaussian variables in the moment space. 
Whenever the Fourier spectrum ${\cal F}_{\rm k}$ has such properties
Eqs.(\ref{61a})-(\ref{62a}) are well defined, and one ought 
to use these also as a criterion for choosing initial conditions.  
Since, however, we are dealing with a time dependent problem, the
question we want to address to ourselves is whether the nonlinear
evolution will distort the asymptotic behavior built into the initial 
state. From the energy viewpoint, we argue that, if the relations
(\ref{629}), (\ref{631}) and (\ref{633}) result a finite energy
density, its conservation will enforce these relations at all
time. 
In fact, as we shall show below, the asymptotic behavior obtained in
(\ref{629}), (\ref{631}) and (\ref{633}) are suffcient to make
(\ref{123a}) finite. 

\medskip
\centerline{\bf B. Asymptotic Conditions and Energy Density}
\medskip

\indent  
In terms of new variables $R_{\rm k}$ and $W_{\rm k}$ the energy
density (\ref{123a}) can be rewritten as
\be
\label{641a}
\frac{\langle H\rangle}{V}\;=\;
\frac{1}{8\pi}\left[\langle\Pi\rangle^2+\langle\phi\rangle^2\right]
-2\int_{\rm k}{\cal E}^{(1)}_{\rm k}
-2\int_{\rm k}{\cal E}^{(2)}_{\rm k} \;\;,
\ee

\noindent where
\bea
\label{641b}
{\cal E}^{(1)}_{\rm k}&=&
\frac{2k}{k_0}\sqrt{k^2_0-R^2_{\rm k}}+\frac{2\bar{m}R_{\rm k}}{k_0}
-\frac{\bar m^2}{k}+\frac{\bar m^4}{4k^2k_0}\;\;,\\\nonumber\\
\label{641c}
{\cal E}^{(2)}_{\rm k}&=&\frac{(\bar{m}-m)k}{k^3_0}W_{\rm k}
             \left(-kR_{\rm k}+m\sqrt{k^2_0-R^2_{\rm k}}\right) \;\;.
\eea

\noindent
Substracting an unimportant constant $2k$ from ${\cal E}^{(1)}_{\rm
k}$ we find, after a little algebra,  
\be
\label{642}
{\cal E}^{(1)}_{\rm k}=
-k\Bigl[\frac{R_{\rm k}}{k_0}-\frac{\bar m}{k}\Bigr]^2
+\frac{k}{k_0}\Bigl[\frac{\bar m^4}{4k^3}
-\frac{R^4_{\rm k}}{k_0(k_0+\sqrt{k^2_0-R^2_{\rm k}})^2}\Bigr]
\ee

\noindent
from which it follows that
\be
\label{643}
\lim_{k \rightarrow \infty} {\cal E}^{(1)}_{\rm k}=
-\frac{k}{k^2_0}\left(-\frac{\bar{m}m^2}{k(k+k_0)}+S_{\rm k}\right)^2
+\frac{k}{k_0}\Bigl[\frac{\bar m^4}{4k^3}
-\frac{\bar m^4}{4k_0^3}
+ {\cal O}\left( \frac{1}{k^4} \right) \Bigr]\;\;.
\ee

\noindent
where we have used (\ref{629}).
It is clear now that this expression decrease as fast as $k^{-5}$ and
$\int_{\rm k} {\cal E}^{(1)}_{\rm k}$ is finite. 
Comparing next (\ref{641c}) with (\ref{628c}) one notes immediately that 
\be
\lim_{k \rightarrow \infty} {\cal E}^{(2)}_{\rm k} 
\;\;\sim \;\;
\lim_{k \rightarrow \infty} {\cal F}^{(2)}_{\rm k} . 
\ee

\noindent
Hence, $\int_{\rm k}{\cal E}^{(2)}_{\rm k}$ is finite when the condition
(\ref{633}) is satisfied. 

\indent
The above qualitative discussion shows that the prescriptions given in
(\ref{629}), (\ref{631}) and (\ref{633}) can make energy density finite. 
They will thus define the initial states and the system will
develop a well defined time evolution within a limited moment space. 
Because of the energy conservation, the asymptotic behavior of the
Gaussian variables essentially will not be modified by the dynamics,
otherwise this will cost infinite amount of energy from the
system. 
This discussion has not considered, however, possibilities of eventual
runaway, caused by eventual unbounded below behavior of the potential. 
The studies of this problem will require further detailed analysis of
$\langle H\rangle$ as multidimensional function of $\bar m$,
$R_{\rm k}$ and $W_{\rm k}$, (one coordinate for each $k$), and
verification of its funcional properties \cite{KL95}.  

\smallskip
\begin{center}
{\bf VI.  Concluding Remarks}
\end{center}
\smallskip

\indent
In summary, we have presented in this paper a framework to treat the
initial-value problem for interacting fermion-scalar field models. 
The method allows one to describe the real-time evolution of the
fields in terms of the dynamics of few observables yielding a set of 
self-consistent equations for expectation values of linear and
bilinear field operators. 
Although the procedure  is quite general, we have however implemented
the calculation within the simplest context of relativistic scalar
plasma system.  
We have shown in detail that the usual form of renormalization also
applies to the present  nonperturbative calculation and we have
obtained finite expression for energy density. 
A simple numerical calculation suggests that the system has always a
single stable minimum, although further investigation will be
necessary for other oscillation modes.  
The standard approach to this question is throught the use of RPA
analysis, where the stability is indicated by its the eigenvalues.   
It is interesting to mention here that the excitation modes described
by the RPA equations are the quantum particles of the field. 
In fact, the physics of one meson and two spin-1/2 fermion can be
investigated from this equation. 
We have also discussed the renormalization for the time-dependent
equations. 
Using energy conservation as the key, we found that there
are a finite range around the vacuum where the dynamics of the system
is well defined. 

\indent
Finally, we comment that systematic corrections to include dynamical
correlation effects to the present mean-field calculations can in
principle be readily applied  with the help of a projection technique
discussed in I and Ref.\cite{TNTP97}.  
In this case, the occupation numbers are no longer constant and will
affect the effective dynamics of the Gaussian observables. 
The framework presented here serves also as groundwork to finite
density and finite temperature discussions \cite{TTP97}. 
In particular, a finite-matter density calculation beyond the
mean-field approximation allows one to study collisional observables
such as transport coefficients \cite{TP87}. 
The extension of this procedure to explore nonuniform systems is
straighforward but lengthy. 
In this case, the spatial dependence of the field are expanded in
natural orbitals of extented one-body density. 
These orbitals can be given in terms of a momentum expansion through
the use of a more general Bogoliubov transformation \cite{RS80}.

\vskip 0.7cm
\centerline{\Large\bf ACKNOWLEDGMENTS}
\vskip 0.7cm

E.R.T.N. and C-Y. L. are supported by Conselho Nacional de
Desenvolvimento Cient{\'{\i}}fico e Tecnol\'ogico (CNPq), Brazil.

\centerline{\Large\bf Figure Captions}

\vskip 0.7cm

FIGURE 1. The behavior of the ground-state mean-field energy density 
$E(x)$ of the scalar plasma system as a function of fermionic 
effective mass $x=g\langle\phi\rangle/m=1-{\bar m}/m$ for any values of 
the coupling constant $g$ and mass scale ${\mu/m}=0.1$ fixed.

FIGURE 2. The behavior of the ground-state mean-field energy density 
$E(x)$ of the scalar plasma system as a function of fermionic 
effective mass $x=g\langle\phi\rangle/m=1-{\bar m}/m$ for any values of 
the coupling constant $g$ and mass scale ${\mu/m}=2$ fixed.

FIGURE 3. The behavior of the ground-state mean-field energy density 
$E(x)$ of the scalar plasma system as a function of fermionic 
effective mass $x=g\langle\phi\rangle/m=1-{\bar m}/m$ for coupling
constant $g^2=\pi/100$ and mass scale ${\mu/m}=2$.

FIGURE 4. The behavior of the ground-state mean-field energy density 
$E(x)$ of the scalar plasma system as a function of fermionic 
effective mass $x=g\langle\phi\rangle/m=1-{\bar m}/m$ for coupling
constant $g^2=\pi/100$ and mass scale ${\mu/m}=0.1$.

\end{document}